\documentclass[%
reprint,
superscriptaddress,
 amsmath,amssymb,
 aps,prx
]{revtex4-2}

\usepackage{graphicx}% Include figure files
\usepackage{dcolumn}% Align table columns on decimal point
\usepackage{bm}% bold math
\usepackage{braket} % for \ket and related commands
\usepackage[colorlinks=true,linkcolor=blue,urlcolor=blue,citecolor=blue]{hyperref}% add hypertext capabilities

\usepackage{ragged2e}
\usepackage{subfigure}
\usepackage{subcaption} 
\usepackage{xcolor}
\usepackage{colortbl}
\usepackage{float}
\usepackage{physics}
\usepackage{float}

\usepackage{graphicx}% Include figure files
\usepackage{dcolumn}%
\usepackage{tikz}
\usepackage{adjustbox}
\usepackage{amsmath, amssymb}
\usepackage{booktabs}
\usepackage{array}
\usepackage{bm}

\newcolumntype{C}[1]{>{\centering\arraybackslash}p{#1}}
\begin{document}

\preprint{APS/123-QED}
\title{Enhancing the Size of Phase-Space States Containing Sub-Planck-Scale Structures via Non-Gaussian Operations}

\author{Arman}
\email{a19rs004@iiserkol.ac.in}
\affiliation{Department of Physical Sciences, Indian Institute of Science Education and Research Kolkata, Mohanpur-741246, West Bengal, India}

\author{Prasanta K. Panigrahi}%
\email{pprasanta@iiserkol.ac.in}
\affiliation{Department of Physical Sciences, Indian Institute of Science Education and Research Kolkata, Mohanpur-741246, West Bengal, India}

\date{\today}

%is prone to decoherence effect as compared to the prosed single mode states . terms: squeezing and displacement.

\date{\today}

\begin{abstract}

(i) We observe a metrological advantage in phase-space sensitivity for photon-added cat and kitten states (KS) over their original forms, due to phase-space broadening from increased amplitude via photon addition, albeit with higher energy cost. (ii) Using accessible non-classical resources, weak squeezing and displacement, we construct a squeezed state (Sq) and two superposed states: the squeezed Schrödinger cat state (SSD) and the symmetrically squeezed state (SS). Their photon-added variants (PASq, PASSD, and PASS) are compared with parity-matched cat and KSs using quantum Fisher information (QFI) and fidelity. The QFI isocontours reveal regimes where KS exhibit high fidelity and large amplitude ($\beta_{\text{KS}}$), enabling their preparation via Gaussian operations and photon addition. Similar regimes are identified for cat states enhanced by squeezing and photon addition, demonstrating improved metrological performance. Moreover, increased amplitude—and thus larger phase-space area—reduces the size of interferometric fringes, enhancing the effectiveness of quantum error-correction in cat codes.

% Non-Gaussian operations, particularly photon additions, are studied on compass and cat states. In investigating the effect of these operations on the states via Fisher information corresponding to the phase-space shift and thereafter, fidelity closer to 1, we realize the importance of these photon addition operations. We further take two types of states, composed of squeezing and superposition of displacement operators with opposite phase-space directions. The parameter space of squeezing, displacement with ($\beta$) amplitude of the KS is seen for equal Fisher information, against the small phase-space shift, between the KS and the proposed states with photon addition operations. Interestingly, average energy remains the same for competing sets of states against Fisher information, indicating within the same energy, large $\beta$ with high fidelity. In further analysis of this locus, it provides a region of high fidelity with KS, leading to an increase in the amplitude ($\beta$). However, realization of such non-Gaussian operations is possible through probabilistic models such as OPA or JC-like interactions as discussed previously in the literature, It is important to look for approximate deterministic preparation of these photon additions. The proposed states without non-Gaussian operation are possible through the two-photon Rabi model and the single-mode Rabi model with two-photon drive.       

\end{abstract}

\maketitle
\section{Introduction}
% The Gaussian distributions, states with positive quadrature distributions, and photon statistics variance larger than the mean photons, such as coherent, squeezed states, and their thermal contour part, are inclined to a more classical nature as compared to the distribution having either a delocalized and negative quadrature distribution, leading to non-Gaussian packets. It becomes ambiguous to characterize the Gaussian distribution from the Fock space expansion as a photon number distribution apart from Poissonian statistics for coherent states. The coherent states are minimum uncertainty wave packets, following equal conjugate quadrature uncertainties, with being Gaussian in the quadrature space. The lasers at low temperature and a particle in a harmonic oscillator, localized to the ground state, are the simplest instance of the coherent state. The squeezed state, with unequal conjugate quadrature variance, positive and Gaussian Wigner distribution, requires two-photon single-mode absorption and emission in presence of strong coherent pump field, a semiclassical and non-linear $\chi^{(2)}$ type interaction, available in the OPO cavity [citation squeezing].  

Gaussian states~\cite{DavidStoler}—such as coherent and squeezed states, including their thermal counterparts—are characterized by positive quadrature distributions and photon number variances larger than or equal to their mean, which lends them a more classical nature compared to states with delocalized or negative Wigner functions. In Fock space, distinguishing Gaussian states based solely on photon number distributions is often ambiguous, except for coherent states exhibiting Poissonian statistics. Coherent states are minimum-uncertainty wave packets with equal variances in conjugate quadratures and Gaussian profiles in phase space. The common realizations include laser light at low temperatures and the ground state of a quantum harmonic oscillator. Squeezed states, on the other hand, exhibit unequal quadrature variances while maintaining a positive Gaussian Wigner distribution. They are typically generated via two-photon, single-mode interactions under a strong coherent pump field, representing a semiclassical nonlinear process governed by $\chi^{(2)}$ interactions, as realized in optical parametric oscillator (OPO) cavities~\cite{Henning_squeezed}.

Non-Gaussian states~\cite{NG-filip, NG-matteio, 11191220} exhibit delocalized quadrature distributions across both positive and negative regions in phase space. Such behavior can arise from diffusive bath interactions \cite{someone2025metrology, BlaisDiffusive} or from coherent, dispersive single-mode nonlinear interactions beyond the $\chi^{(2)}$ regime. The Kerr-type interactions, theoretically predicted \cite{lakshmibala2022revivals} and experimentally demonstrated \cite{kirchmair, shruti-puri}, disperse initially Gaussian states in phase-space, generating negative Wigner regions. This evolution leads to highly non-Gaussian, non-classical states such as cat and compass states, which exhibit interference fringes with sub-Planck-scale structures \cite{Zurek2001}. The size of these fringes, centered around the origin, is inversely proportional to the state's phase-space support (action), making them highly sensitive to small displacements ($D[\lambda_{s}]$) and phase rotations ($e^{i\lambda_{\phi}\hat{n}}$). These states can locally saturate the quantum Cramér–Rao bound \cite{cavesgeometry}, achieving the Heisenberg limit with estimation precision $\delta\lambda_{s}$ (or $\delta\lambda_{\phi}$) scaling as $1/\sqrt{\langle\hat{n}\rangle}$ (or $1/\langle\hat{n}\rangle$) in the limit $\lambda_{s}, \lambda_{\phi} \rightarrow 0$ \cite{Toscano}, underscoring their metrological relevance. Furthermore, as approximate eigenstates of the annihilation operator $\hat{a}$, these states are inherently robust against photon loss, forming the basis of error-correcting schemes such as the cat code \cite{cat_code0,cat_code1}.

The generation of highly non-classical states typically requires either strong non-linear interactions—such as the Kerr-type nonlinearity characterized by the term \(\hat{a}^{\dagger\,2}\hat{a}^{2}\)—or access to atom-field entanglement~\cite{hacker2019deterministic} within the strong dispersive regime~\cite{hastrup-GKP}, which has been experimentally demonstrated~\cite{bild2023schrodinger}. Jaynes-Cummings (JC) type interactions, which are central to this process, are accessible in both optical~\cite{agarwal-damp} and superconducting circuit platforms~\cite{shruti-puri}. In trapped atoms and ions, the strong coupling strength enables the formation of non-Gaussian states in their motional degrees of freedom, which are intricately linked to their vibrational modes~\cite{hastrup2021measurement}. Experimental realizations of non-classical states, such as Schrödinger cat and compass states, have been achieved in these atomic and ionic systems. Another promising platform is optomechanics, where the optical field couples to a mechanical membrane. In such systems, the position modes of the mechanical element can be driven into non-Gaussian states through an interaction that depends on the optical field intensity ($\langle\hat{n}\rangle$). The cat and compass states in the mechanical modes have also been demonstrated in these optomechanical setups~\cite{Quantum_Hypercube_States}.

In addition to the unitary evolution governed by non-linear Hamiltonians and exact interactions—which can intrinsically stabilize cat and compass states as part of their eigenstate spectra—another approach involves multi-mode interactions, particularly of the system-ancilla type. In this framework, engineered interactions are applied either in a single step or through multiple iterations, followed by conditional measurements on the ancilla subsystem. These measurements effectively steer the primary system into a desired non-Gaussian regime~\cite{hastrupnature}. By carefully optimizing the strength and structure of the interactions, and repeating the measurement protocol, it becomes possible to realize an unconditional preparation of the target non-classical states~\cite{Squeezed_hastrup}.

Apart from the unitary evolution from non-linear Hamiltonian and exact interactions, which can stabilize cat and compass states in one of their eigenstates' spectra, there are multi-modal type interactions (system-ancilla type), leading to the engineering of such states via either single or multi-step application of operations. Thereafter, conditional measurement on the ancilla party is used to drive the system in the regime of interest, i.e., to the non-Gaussian distribution. The repeated applications of such measurements with specifically optimized interactions' strengths may lead to the unconditional preparation of the desired states~\cite{Squeezed_hastrup}. 

The possible and popular operations to introduce non-Gaussianity~\cite{NG0, NG1, NG2} in a small amount in the Gaussian~\cite{tara-ag, Asoka-ag}, as well as a non-Gaussian distribution~\cite{Arman}, are photon additions, subtractions, and catalysis and require conditional measurements, which degrade their success probabilities. Maximally observed photon addition is in the form of two-photon subtraction in experiments aimed at generating large-amplitude cat states \cite{prl08}. Photon subtraction can be achieved through low transmissivity of the beam splitter (BS) \cite{Asoka-ag,BAEVA1,baeva2,kim,chen2023generation,podoshvedov2019,podoshvedov2023}, allowing for the expansion of the unitary in the linear form, while photon addition is possible with a $\chi^{(2)}$ nonlinear crystal (BBO) in optical platforms \cite{Zavatta}, also known as optical parametric
amplifiers (OPA). This nonlinearity is also feasible in the circuit QED. Moreover, JC interactions in the weak coupling regime can employ entangling operations to achieve photon addition and subtraction, typically disregarding higher-order contributions from unitary expansion. These methods are successful in implementing photon addition; however, their efficiency (success probability times fidelity) tends to decrease with repeated light applications.   

 Photon addition results in an increase in the average photon number of the state as well as its non-Gaussian nature \cite{tara-ag, Zavatta, Arman}. The presence of non-Gaussianity describes the non-classical features of the state. In this paper, we consider different non-classical states: cat, compass (KS), superposed squeezed and displaced states, which have significant resemblance in their phase-space distribution \cite{armancompass}, enhancing non-classical nature by adding photons, hence their photon-added variants~\cite{Arman,akhtar0} (or subtracting photons~\cite{Asoka-ag}) and potentially increasing their amplitude. These states, with photon addition, exhibit similar sensitivity to the cat and KS but with increased amplitude. We use fidelity and Fisher information metric to show quantitative relative analysis for the proposed and target states. Having the same Fisher information only indicates identical sensitivity to external perturbations or small phase-space shifts, while fidelity relates the proximity of these states to the target (cat state and KS).

However, this work focuses mainly on photon additions; operations such as additions and subtractions are non-unitary, can not be realized through single-mode party evolutions without ancilla, challenges to their experimental realization. These operations can possibly be represented as a superposition of the unitaries, which requires again JC, Rabi, and strong dispersive interaction regimes, which may increase the success probabilities of non-Gaussian operations.

In this paper, we study the contour of quantum fisher information and fidelity (Sec.~\hyperref[sec:III]{IV}) for two states proposed in Sec.~\hyperref[sec:II]{III}, focusing on the close resemblance quantitatively with KS and cat states in terms of both squeezing ($r$) and displacement ($\alpha$) parameters. Finally, Sec.\hyperref[sec:VII]{V} contains a conclusion with a summary.

 % The squeezing is not limited to its uses in a single mode, as recently shown for two-mode squeezing superposition \cite{twomode} with metrological aspects and ion trap implementation. Additionally, we analyze the fidelity between the proposed states and the compass state in Sec.\hyperref[sec:IV]{IV}, and propose a theoretical model for their preparations in Sec.\hyperref[sec:V]{V}.

% In this paper, we study the phase-space structures (Sec.\hyperref[sec:III]{III}) for two states that are proposed in Sec.\hyperref[sec:II]{II}. The probability number distribution (PND) is studied for both squeezing and displacement parameters of the state. The squeezing is not limited to its uses in single-mode as is shown recently for two-mode squeezing superposition \cite{twomode} with metrological aspect and their ion trap implementation. Further, with the help of fidelity between proposed states and compass state, we investigate their similarities and sensitivity in the Sec.\hyperref[sec:IV]{IV} and hence, a theoretical model for their preparations in Sec.\hyperref[sec:V]{V}. Finally, Sec.\hyperref[sec:VII]{VII} contains a conclusion with the summary. 

\section{Interference fringes and effect of photon additions}
\begin{figure}[htbp]
\centering
\includegraphics[height=0.5\linewidth, width=\linewidth]{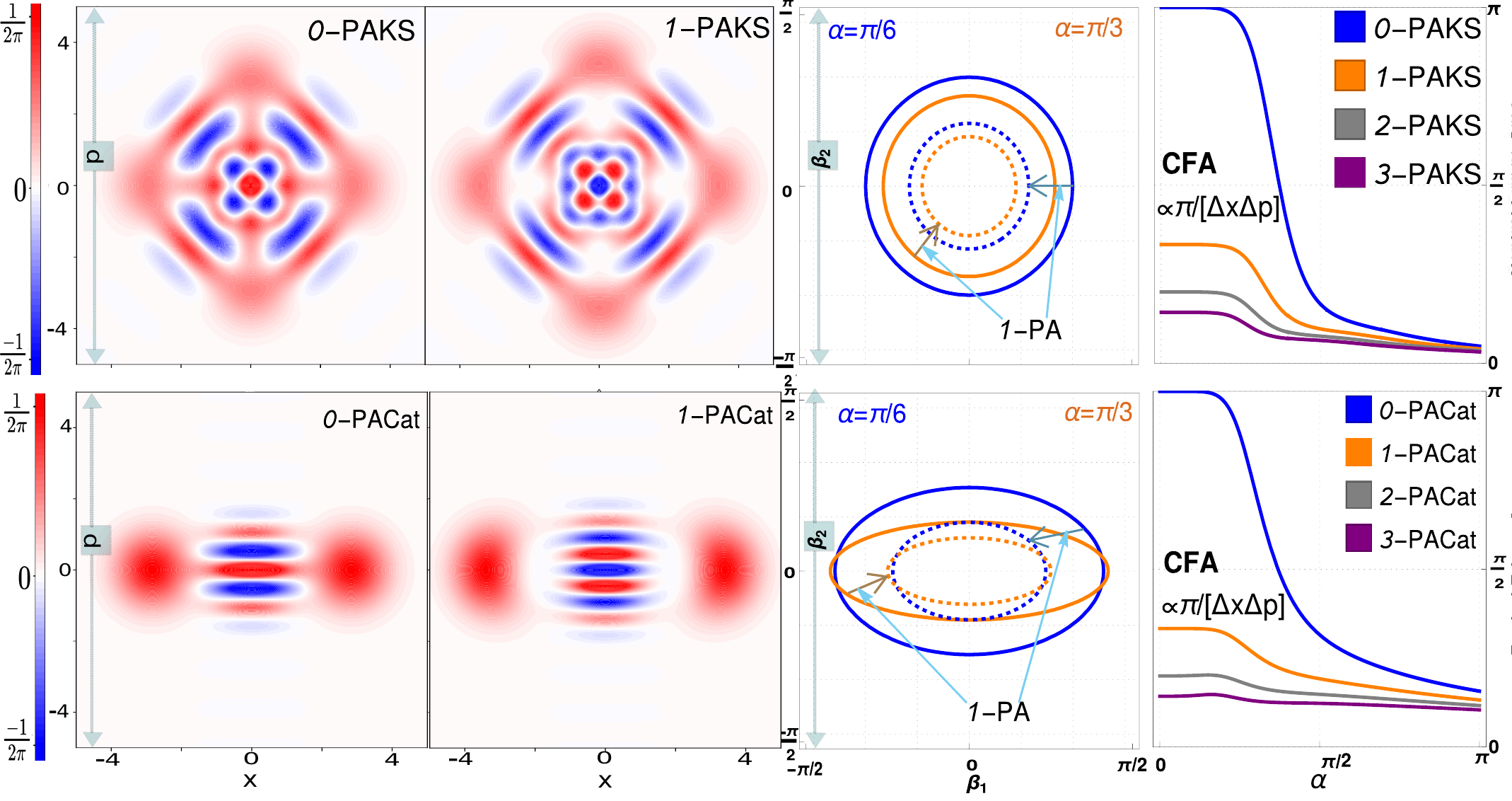}
 \caption{\justifying (Color online) Evident from phase space distribution (first two left column) and central fringe area (CFA), there is an indication of enhancement for phase-space area (support) and metrological power (in $3^{rd}$ column for $O_{\lambda_{o}}$) for both cat and compass state upon photon addition. The relation is clearly seen in the rightmost column for CFA between adding photons and increasing amplitude $(\alpha)$ for the cat state and KS.} 
 %Effect of photon additions  Wigner distribution on first two QFI ($F_{Q}$) and fidelity ($\mathcal{F}$) comparison between photon-added Kitten States (PAKS) and their corresponding parity-matched Kitten State (KS) counterparts.
    \label{fig:KSCatillust}
\end{figure}

This section illustrates effect of photon additions on the cat and compass states via measures Wigner function $\left(W_{\beta}=(2/\pi)\Tr[\rho\,\hat{\Pi}\,\hat{D}[2\beta]], \text{with}\, \hat{\Pi}=e^{i\pi \hat{n}}\, \text{and}\, \hat{D}[2\beta]=\right.$\\$\left.e^{2\beta \hat{a}^{\dagger}-2\beta^{*}\hat{a}}\,\text{is displacement operator} \right)$, overlap of such phase-space distribution $\left( O_{\lambda}= 2\,(\pi^{-1})\right.$ $\left.\int_{-\infty}^{\infty}d^{2}\beta\, W_{\beta}W_{\beta + \lambda}\right)$ and central fringe area (CFA) of same overlap (fringes consequent from Wigner distribution) which quantifies sensitivity of both states against small phase-space displacements ($\lambda$). The change in parity and increase in spread or support of phase-space distribution appears for both KS (upper row) and cat state (lower row) upon single photon addition from 1st to 2nd column, in the Fig.\ref{fig:KSCatillust}. Further, locus of first zeros ($\lambda_{o}$) for overlap $\left( O_{\lambda_{o}}=1-|\lambda_{o}|^{2}F_{Q}^{\psi}(\theta)/4 \right.$, being true for $\left.|\lambda_{o}|<<1\right)$ for both state, approaches to phase space origin upon PA operations as seen in Fig.\ref{fig:KSCatillust} $3^{rd}$ column, while maintaining same original amplitude ($\alpha$). This shows decreasing central fringe size of Wigner distribution and overlap with rise in their sensitivity to smaller shifts upon PA as compared to no PA. Similar behavior is seen with increasing those states' amplitude ($\alpha$), leading to a relation with the PA operations. The decreasing area is further evident in the rightmost column in Fig.\ref{fig:KSCatillust} for subsequent PAs. This leads to a complete illustration of the effect of photon additions in phase space and quantitative measures, such as the fringe size of the KS and Cat states. 
In this paper, we use some symbols and notation as $\hat{n}=\hat{a}^{\dagger}\hat{a}$, the number operator with $\hat{a}\,(\hat{a}^{\dagger})$ being the bosonic annihilation (bosonic creation) operator. The $\hat{a}^{\dagger}$ represents photon addition (PA), while $\hat{a}$ represents photon subtraction (PS).
Similarly, $W_{1}$ metric~\cite{laxmi-paul} and entropy-based measure, important in quantum thermodynamics, such as KL-divergence related to $F^{\psi}_{Q}$, can also capture operations such as PA.
% We will discuss the phase-space structure of the cat and compass state with and without photon addition to qualitatively investigate fringe size, as well as quantitatively overlap and its second-order derivative (analogous to speed limits). This second-order derivative with respect to the parameter $\theta$, either displacement or rotation, is quantum Fisher information. Interestingly, Fisher Information can be related similarly again to the entropy-type measure, which is symmetric KL divergence. This divergence is the separation between the two distributions, caused from the random occurrence of the variable.
% In this paper, we use some symbols and notation as $\hat{n}=\hat{a}^{\dagger}\hat{a}$, the number operator with $\hat{a}\,(\hat{a}^{\dagger})$ being the bosonic annihilation (bosonic creation) operator.
 
\section{Photon addition on superposition of squeezed and displaced states}\label{sec:II}

Our primary goal is to enhance the effective amplitudes of small cat and compass states using non-Gaussian operations. In particular, photon addition (PA) and photon subtraction (PS) modify the mean energy of an initial state $\ket{\psi}$ according to  
\begin{align*}
\langle \hat{n} \rangle_{\text{PA}}
&= \frac{\langle\Delta^{2}\hat{n}\rangle_{\psi}}{\langle\hat{n}\rangle_{\psi}+1}+\langle\hat{n}\rangle_{\psi}+1, \\
\langle \hat{n} \rangle_{\text{PS}}
&=\frac{\langle\Delta^{2}\hat{n}\rangle_{\psi}}{\langle\hat{n}\rangle_{\psi}}+\langle\hat{n}\rangle_{\psi}-1.
\end{align*}
These relations show that PA increases the available energy resource and therefore can enhance the effective amplitude of the state, whereas PS does not necessarily do so, significantly true for the sub-Poissonian light.

This effect is particularly pronounced for cat and kitten (KS) states and their higher superpositions, whose mean photon number satisfies  
\begin{equation*}
\langle \hat n \rangle_{KS} = \mathbb{F}(\left|\beta_{KS}\right|),
\end{equation*}
where $\mathbb{F}$ is a monotonic and increasing function of the coherent amplitude $|\beta_{KS}|$ (see $2^{nd}$, $4^{th}$ and $5^{th}$ row in Table.\ref{tab:proposed_state}). Although PA $(\hat{a}^\dagger)$ and PS $(\hat{a})$ act differently on the state, they cannot be distinguished by parity $\hat{\Pi}$ measurements alone for cat and KS~\cite{Arman}, since both operations shift the parity by $\pi$,
\begin{equation*}
\hat{\Pi}=e^{i\pi \hat{n}}
\;\xrightarrow[\;\text{on PA/PS}\;]{\;\text{shift by}\,\pi\;}\;
e^{i\pi(\hat{n}+1)}.
\end{equation*}
% However, the mean photon number clearly differentiates the two operations: for $\psi\in\{\mathrm{Cat},\mathrm{KS}\}$, photon addition increases the effective coherent amplitude $\beta$, whereas photon subtraction leaves $\beta_{\psi}$ unchanged or slightly affected as to PA case. This follows from the relation $\langle \hat{n} \rangle_\psi=\mathbb{F}(|\beta_{\psi}|)$ and from the fact that these states are eigenstates of powers of the annihilation operator $\hat \hat{a}^k$ ($k\in\mathbb{N}$) relevant to PS (making shuffling only in elements of basis of either Cat or KS, respectively leaving amplitude $\beta_{\psi}$ unchanged upon PS where $\psi \in \{\mathrm{Cat},\,\mathrm{KS}\}$). 

However, the mean photon number, apart from Wigner function and its negativity, clearly distinguishes the two operations: for 
$\psi\in\{\mathrm{Cat},\mathrm{KS}\}$, photon addition increases the effective coherent amplitude $\beta_{\psi}$, whereas photon subtraction leaves $\beta_{\psi}$ unchanged, or only weakly modified in comparison with the PA case. This follows from the relation 
$\langle \hat{n} \rangle_{\psi}=\mathbb{F}(|\beta_{\psi}|)$ and from the fact that these states are eigenstates of powers of the annihilation operator $\hat{a}^{k}$ ($k\in\mathbb{N}$) relevant to PS. As a consequence, photon subtraction primarily induces a reshuffling among the basis components of either the Cat or KS, thereby leaving the amplitude $\beta_{\psi}$ effectively unchanged for $\psi\in\{\mathrm{Cat},\mathrm{KS}\}$.

Consequently, photon-addition provides a more efficient route for amplifying the size of highly non-classical states with sub-Planck-scale phase-space structures. To quantify this enhancement, we compare the overlaps of cat and KS of the same parity with their photon-added and photon-subtracted counterparts, following Refs.~\cite{armancompass, cat_code1}.

%It is easy to show that above states shift their parity $\left(\hat{\Pi}=(-1)^{\hat{n}}\right)$ by $\pi$ due to PA and PS operations individually, leading to a jump in these states' orthogonal basis, while PA increase average energy as well as amplitude ($\beta$) and PS does not affect amplitude ($\beta$) because of being eigenstate of the annihilation operator ($\hat{a}^{k}, \forall\, k\in \mathbb{N}$).

% Further, we lay out one of our motivation for using PA operation in enhancing the cat and KS operations. Similar, investigation has been performed in the previous literature~\cite{armancompass} for KS, detailing analysis over increasing phase-space area of KS via tuning parameters of squeezing, displacement and Fock states. The role of Fock state ($\ket{n}$) is similarly played by the PAs in this paper. On the same footing, refs.~\cite{Arman,chen2023generation} discusses and gives hint for enhancing amplitude for the cat state.  

Furthermore, we present one of our key motivations for employing the PA operations to enhance cat and kitten states. A related investigation was carried out in Ref.~\cite{armancompass}, where the phase-space area of kitten states was enlarged by tuning parameters such as squeezing, displacement, and the incorporation of Fock states ($\ket{n}$). In the present work, the role played by Fock states ($\ket{n}$) in that study is effectively realized through successive photon additions. Along similar lines, Refs.~\cite{Arman, chen2023generation} also explore, and suggest the possibility of, amplitude enhancement in cat states via PAs. 

% In the previous work~\cite{armancompass}, it was shown that the amplitude of KS is highly affected by the operations of the squeeze, displacement, and Fock states. With increase in the Fock state $(\ket{n})$ for the above operations of squeezing and displacement, has allowed to have large amplitude of the KS with large overlap as $n$ increases because the amplitude and mean photon number is correlated for the KS. Although this is not direct intuitive to think that KS and SSD, and SS are directly related to each other, even if these had the same mean energy. Our intuition started from the phase-space distribution and became justified when corresponding properties are studied, and were almost similar for PND, sensitivity as quantum Fisher information (QFI), finally finding the region of high fidelity.
% These cat and compass states have been discussed in the context of application in quantum metrology in optical as well as trapped ion platforms, quantum error correction in available computing platforms such as microwave, and communication in the optical platform, as discussed in the previous work \cite{armancompass}.

We start with the three proposed states, which are important candidates for increasing the amplitudes of the cat and KS via non-Gaussian operations (PAs). Two out of these have been discussed in the previous work~\cite{armancompass} in the practical system and the introduction, are squeezed superposed displaced state (SSD) and superposed squeezing state (SS), while remaining is the squeeze state (Sq), a widely used method in preparing cat states. These states can be prepared in the superconducting circuit and trapped ions with high fidelity and high success probability, while with low success probability in the optical platform. The PA variants of these proposed states, $n$-photon-added SSD ($n-$PASSD), $n$-photon-added SS ($n-$PASS), and $n-$photon-added squeeze state ($n-$PASq), are compared with target states, KSs, and cat states shown in the Table.~\ref{tab:proposed_state} as well as in the Appendix~\ref{appendix:b}. 

For clarity, we introduce a compact labeling scheme for pairs of target and proposed states used in the quantitative analysis. Specifically, we define 
\textbf{prstrg-\#1} for the pair 
$\big[\psi_{n\text{-PASSD}} \;\text{vs.}\; \psi^{l}_{\mathrm{KS}}(-)\big]$, 
\textbf{prstrg-\#2} for 
$\big[\psi_{n\text{-PASS}} \;\text{vs.}\; \psi^{l}_{\mathrm{KS}}(+)\big]$, 
and \textbf{prstrg-\#3} for 
$\big[\psi_{n\text{-PASq}} \;\text{vs.}\; \psi^{l}_{\mathrm{cat}}\big]$. 
These pair labels are employed throughout Sec.~\ref{sec:III}. In addition, an analogous analysis is carried out for pairs consisting exclusively of target states. We label these as \textbf{trgtrgn-\#1} for 
$\big[\psi^{l}_{\mathrm{KS}}(-) \;\text{vs.}\; \hat{a}^{\dagger n}\psi^{l}_{\mathrm{KS}}(-)\big]$, 
\textbf{trgtrgp-\#2} for 
$\big[\psi^{l}_{\mathrm{KS}}(+) \;\text{vs.}\; \hat{a}^{\dagger n}\psi^{l}_{\mathrm{KS}}(+)\big]$, 
and \textbf{trgtrgE-\#3} for 
$\big[\psi^{l}_{\mathrm{Cat}} \;\text{vs.}\; \hat{a}^{\dagger n}\psi^{l}_{\mathrm{Cat}}\big]$.
Here, $\bar{\mathsf{f}}_{k_{1},k_{0}}=2^{k_{0}/2}\left[\sin{\left(\frac{(2k_{1}+1)\pi}{4}\right)}\right]^{k_{0}}$ with $k_{0}=0$ denotes state $\psi_{ks}^{l}(+),$ while value $k_{0}=1$ relates to $\psi_{KS}^{l}(-).$ For the sake of the reader, we have included both these KSs in the Table.~\ref{tab:proposed_state}, having same form as in the ref.~\cite{armancompass} and all the acronm. are available in Appendix~\ref{appendix:a}.

In the context of quantum metrology, Cat, Sq, KSs, SSD, and SS are useful in the precision measurement of small displacements, effectively resulting from either coherent drive or bath noise. These small effective displacements are represented as unitary displacement operators $D[s]$. The magnitude of these small displacements ($|s|$) can be estimated with precision—identified via the sensitivity measure—the QFI ($F_{Q}^{\psi}$)~\cite{liuquantum, cavesgeometry} for probe $\psi$, associated with the Cramer-Rao bound. In case of pure states $\psi$, $F_{Q}^{\psi} = 4\langle\Delta^{2}\hat{x}(\theta)\rangle$ (Variance) where generator $\hat{x}(\theta)=\left.iD^{\dagger}[|s|e^{i\theta}]\partial_{|s|}D[|s|e^{i\theta}]\right|_{s\rightarrow 0}$~\cite{Zurek2001, armancompass}. The higher the Fisher information of a given probe state, the greater its sensitivity to small changes in the parameter affecting that state. In other words, this measure can distinguish parameter-dependent density matrices from neighboring ones that exist in the parameter space. Therefore, it is pertinent, apart from the high fidelity regime, to demonstrate that photon-added variants ($n-$PAKS and $n-$PACat) exhibit the same metrological power $(F_{Q}^{\psi})$ for Cat and KSs $w.r.t$ parameter variations—particularly for displacements in phase space—although other types of parameter-dependent processes also exist, such as phase shifts and decoherence~\cite{agarwal-damp, armancompass}. This analysis may manifest our motivation for PAs on proposed states if there are high-fidelity regimes for pairs \textbf{trgtrgn-\#1}, \textbf{trgtrgp-\#2}, and \textbf{trgtrgE-\#3}.

% Before making a jump to the study of QFI and thereafter, finding closeness of proposed and target (KS and Cat state) via fidelity $(\mathcal{F})$ measure, we take the same investigation among KS, Cat state and its PA variants ($n-$PAKS) to explore that this PA type non-Gaussianity leads to high fidelity and large amplitude regions for both KS and Cat state. Considering the two KSs' combinations, shown in the Table.\ref{tab:proposed_state}, form two pairs with $n-$PAKS as consisting of negative labeled KS, namely pair \textbf{trgtrgn$-\#1$} [$\psi_{ks}^{l}(-)$ and $\hat{a}^{\dagger\,n}\psi_{ks}^{l}(-)$] and positive labeled KS, namely pair \textbf{trgtrgp$-\#2$} [$\psi_{ks}^{l}(+)$ and $\hat{a}^{n\,\dagger}\psi_{ks}^{l}(+)$].

\begin{figure*}[htbp]
\centering
\includegraphics[width=\linewidth]{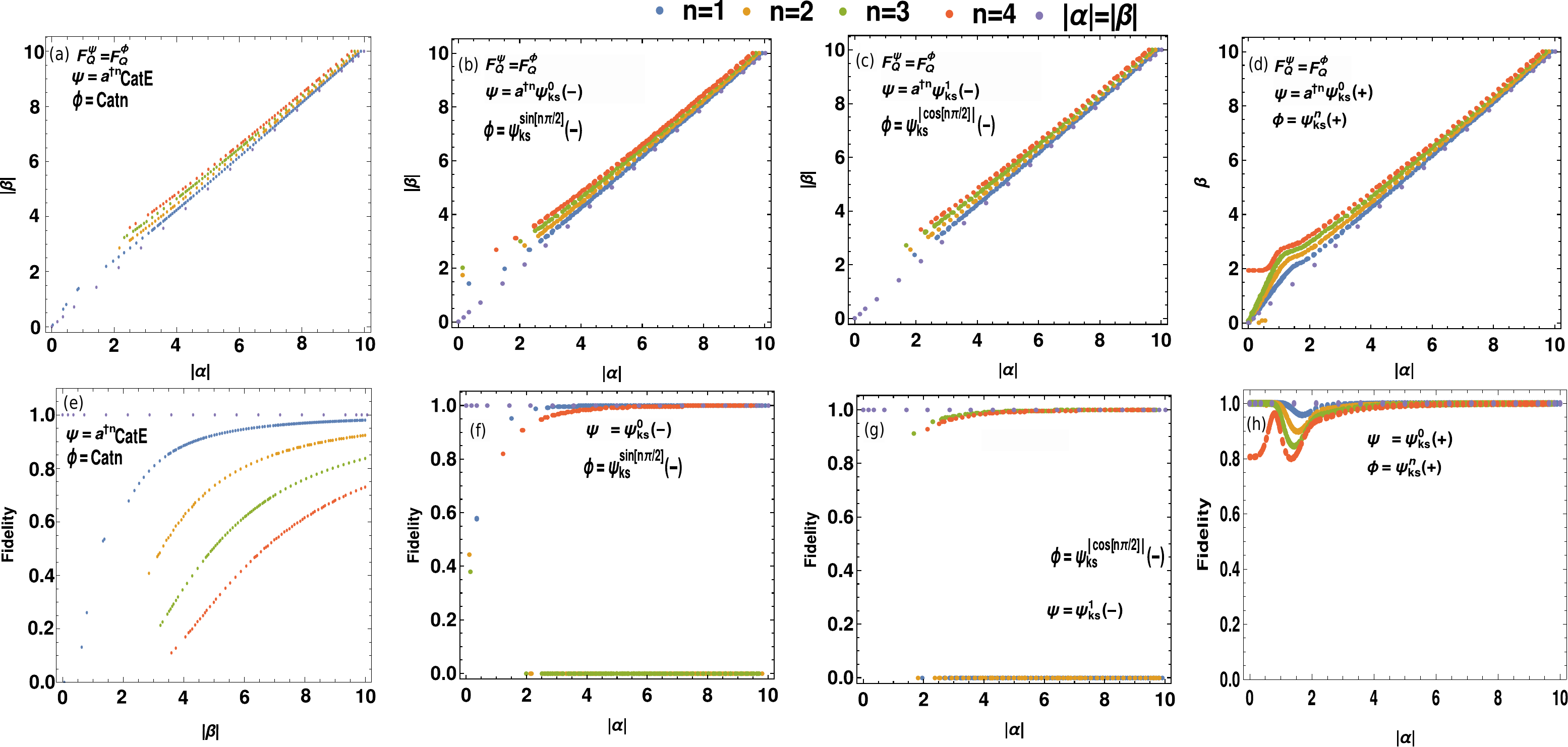}
 \caption{\justifying (Color online) QFI and fidelity ($\mathcal{F}$) comparison among $n-$PACatE, $n-$PAKS, and their corresponding parity-matched CatE and KS counterparts. The top panels (a–d) illustrate QFI constraints in space of coherent amplitudes $|\alpha|\in \hat{a}^{\dagger\,n}\psi$  and $|\beta|\in\psi$. Specifically: (a) takes pair $\{\hat{a}^{\dagger\,n}\text{CatE},\psi_{\text{Cat}}^{n}\}$; (b) compares $\hat{a}^{\dagger\,n}\psi_{KS}^{0}(-)$ with $\psi_{KS}^{l}(-)$ where $l = \sin(n\pi/2)$; (c) contrasts $\hat{a}^{\dagger n}\psi_{KS}^{1}(-)$ with $\psi_{KS}^{l}(-)$ where $l = |\cos(n\pi/2)|$; and (d) identifies a QFI equality condition between $\hat{a}^{\dagger n}\psi_{KS}^{0}(+)$ and $\psi_{KS}^{n}(+)$. The lower panels (e-h) present fidelity across the same parameter space, showing large $\mathcal{F}$ for amplitudes as a consequence of multiple $n$ photon addition, shifting the equal QFI contours to enhanced amplitudes i.e., $|\beta|>|\alpha|$ .}
    \label{fig:all_images}
\end{figure*}

In our investigation, the competing states in each pair are made such that the parity $(\hat{\Pi})$ remains the same. Additionally, each state $\psi\in \{\psi_{KS}^{l}(-),\, \psi_{KS}^{l}(+),\, \psi^{l}_{Cat}\}$ occupies a distinct element of an orthogonal basis, designated by $l\in \mathbb{N}$. After photon-addition (PA) to any state belonging to specific $l$ within the same basis leads to the state in same orthogonal subspace of parity i.e. $\{\psi^{l},\hat{a}^{\dagger\,n}\psi^{l}\}$ have same parity. Specifically, the orthogonal basis $\{\psi\}$ corresponding to either $\psi_{KS}^{l}(-)$ or $\psi_{Cat}^{l}(-)$ consists of two states labeled by $l \in \{0,1\}$, while the basis associated with $\psi_{KS}^{l}(+)$ comprises four states labeled by $l \in \{0,1,2,3\}$. Each pair has a condition for $l$ relating to the $n$, follows the same parity within the pair (see Fig.\ref{fig:all_images}). The evaluation of QFI and, thereafter, substituting in the constraint $F_{Q}^{\text{KS}}=F_{Q}^{\text{$n-$PAKS}}$ leads to the loci $(\alpha, |\beta|)$ for pair \textbf{trgtrgn-\#1} in Fig.\ref{fig:all_images}(b-c), revealing $|\beta|>\alpha$. Similarly, both pairs \textbf{trgtrgp-\#2} in Fig.\ref{fig:all_images}(d) with constraint $F_{Q}^{\text{KS}}=F_{Q}^{\text{$n-$PAKS}}$ and \textbf{trgtrgE-\#3} in Fig.\ref{fig:all_images}(a) with constraint $F_{Q}^{\text{CatE}}=F_{Q}^{\text{$n-$PACatE}}$, shows the same behavior of $|\beta|>\alpha$ with rising $n$ PAs. Here, $\alpha$ and $\beta$ belong either to KS and PAKS or CatE and $n-$PACatE respectively, in each pair, with $\beta=|\beta|e^{i\pi/4}$ true for Fig.\ref{fig:all_images}(b-c) and $\beta=|\beta|$ true for Fig.\ref{fig:all_images}(a) and (d). In the same fashion, the fidelity $\left(\mathcal{F}=\text{Tr}[\sqrt{\sqrt{\rho_{i}}\rho_{f}\sqrt{\rho_{i}}}]\right)$ analysis is extended for pairs \textbf{trgtrgn-\#1}, \textbf{trgtrgp-\#2} and \textbf{trgtrgE-\#3}. Each pair has $\mathcal{F}$ following the contour for aforementioned constraints of QFI in the respective Figs.\ref{fig:all_images}(a-d). The Fig.\ref{fig:all_images} (a), (b), (c) and (d) revealing loci $(\alpha,\beta)$, show regions of large fidelity in Fig.\ref{fig:all_images} (e), (f), (g) and (h), respectively. This fidelity analysis explores that PAs to the KS or CatE (see table. \ref{tab:nomenclature}) not only reaches same metrological power regimes, but also have larger overlap regions, which necessarily justifies that multiple PAs lead to enlarging of phase-space area for cat and compass states. This motivates us to analyze these same non-Gaussian operations on the SSD and SS.

\begin{table}[h!]
\caption{\label{tab:proposed_state} 
\justifying Main list of analytical expression for proposed states, constructed via the displacement $D[\alpha]$ and squeeze $S[re^{i\phi}]=e^{\frac{r}{2}\,\left(e^{-i\phi}\hat{a}^{2}-e^{i\phi}\hat{a}^{\dagger\,2}\right)}$ operators alongwith photon addition (PA) and target state such as KS and Cat state. Address table.~\ref{tab:nomenclature} for compressive labeling and listing.}
\centering
\rowcolors{2}{gray!10}{white}  % Alternate rows (optional)
\renewcommand{\arraystretch}{1.2}
\begin{adjustbox}{width=1\columnwidth}
\begin{tabular}{C{0.4\linewidth}C{0.55\linewidth}}
\hline
\hline
\rowcolor{gray!20}
\multicolumn{1}{c}{$\mathbf{\textbf{State~} \ket{\psi}}$} & \textbf{State Expression} \\
\hline
$\ket{\psi_{\scalebox{0.5}{n\text{-PASSD}}}}$ &
$\hat{a}^{\dagger n} S[r] \left(D[\alpha] + D[-\alpha]\right) \ket{0}$ \\
$\ket{\psi_{\text{KS}}^{l}(-)}$ &
$\displaystyle \sum_{k=0}^{3} \bar{\mathsf{f}}_{k,1} e^{-i l k \pi / 2} \ket{e^{i k \pi / 2} \beta}$ \\
$\ket{\psi_{\scalebox{0.5}{n\text{-PASS}}}}$ &
$\hat{a}^{\dagger n} \left(S[r] + S[-r]\right) \ket{0}$ \\
$\ket{\psi_{\text{KS}}^{l}(+)}$ &
$\displaystyle \sum_{k=0}^{3} \bar{\mathsf{f}}_{k,0} e^{-i l k \pi / 2} \ket{e^{i k \pi / 2} \beta}$ \\
$\ket{\psi_{\text{cat}}^{l}}$ &
$\displaystyle \sum_{k=0}^{1} e^{i k l\pi} \ket{e^{i k \pi} \beta}$\\
$\ket{\psi_{n\text{-PASq}}}$ &
$\hat{a}^{\dagger\,n}S[r]\ket{0}$\\
\hline
\hline
\end{tabular}
\end{adjustbox}
\end{table}

% Photon-added operations are one of the ways to increase the amplitude of various KSs. We considered two types of KSs' combinations, shown in the table.\ref{tab:proposed_state}. The QFI is evaluated for both KSs and their $n$-photon added variants ($n$-PAKS). Obtaining regions of equal QFI for states $\psi_{ks}^{l}$ against $\hat{a}^{\dagger n}\psi_{ks}^{l}$ ($n$-PAKS) in the Fig.\ref{fig:all_images}(a-c), describes rise in the $\beta$ output amplitude with additions of multiple photons. This comparative analysis involves the same parity of KS and $n$-PAKS for each case, seen in the Fig.\ref{fig:all_images}(a-c). The same set of states is used for the case of fidelity ($\mathcal{F}$), seen in the Fig.\ref{fig:all_images}. Here, the region of fidelity is the locus ($\alpha, \beta$) of equal QFI between a set of states considered in respective Fig.\ref{fig:all_images}(d-f). The respective set of states used in the QFI analysis in the Fig.1 (a), (b), and (c) is the same for fidelity analysis in Fig.1 (d), (e), and (f), respectively. This fidelity analysis shows that photon additions to the KS not only reaches equal QFI regions, there are also larger overlap regions, which necessarily proves that photon addition operations leads to enlarging of phase-space for compass states. This motivates us to analyze these same non-Gaussian operations on the SSD and SS. However, it remains to check these operations effect on the Cat states.

\onecolumngrid

\twocolumngrid

\section{Closeness via Metrological Power ($F_{Q}^{\psi}$)}\label{sec:III}

% {\textcolor{red}{Making comments for $\psi_{1}$ and $\psi_{2}$ about their two-step preparation without non-Gaussian and availability of platforms with their non-Gaussianity ($\hat{a}^{\dagger}$).}} 

% In the context of quantum metrology, KSs and SSD with SS are useful in the precision measurement of small displacement external to the system due to either coherent drive or bath noise. These small displacements are represented as unitary displacement operators $\hat{D}[s]$. The strength of these small displacements can be measured with precision as a result of the sensitivity measure, QFI $(F_{Q}^{\psi})$, relating to the Cramer-Rao bound. 
% As shown in Ref. \cite{armancompass}, the state $\psi_{ks}^{l}(+)$ exhibits no $\theta-$dependence in its QFI $(F_{Q}^{\psi})$. In case of pure states $\psi,$ $F_{Q}^{\psi} = 4\langle\Delta^{2}\hat{x}(\theta)\rangle$ (variance) where generator $\hat{x}(\theta) = iD^{\dagger}[s]\partial_{|s|}D[s]$ [5,27].
% The higher the Fisher information for the considered state as a probe, the more its sensitivity will be to the parameter against which the probe is affected.   

As shown in Ref.~\cite{armancompass}, the state $\psi^{l}_{\text{KS}}(+)$ exhibits no $\theta$-dependence in its QFI ($F_Q^{\psi}$), whereas $\psi^{l}_{\text{KS}}(-)$ and $\psi^{l}_{\text{Cat}}$~\cite{Arman} demonstrate variation with respect to the angular orientation $\theta$ of the Wigner function $W(x,\,p)$. A similar contrast appears in phase space between states SS (or $n-$PASS)~\cite{akhtar0}, SSD (or $n-$PASSD) and Sq (or $n-$PASq), corresponding to $\psi^{l}_{\text{KS}}(+)$, $\psi^{l}_{\text{KS}}(-)$ and $\psi^{l}_{\text{Cat}}(-)$, respectively. Explicit formula for QFI ($\theta$), due to this, we have a different form of constraint presented by $F_{Q}^{\psi}.$

\begin{center}
    \begin{figure}[H]
    \includegraphics[width=\linewidth]{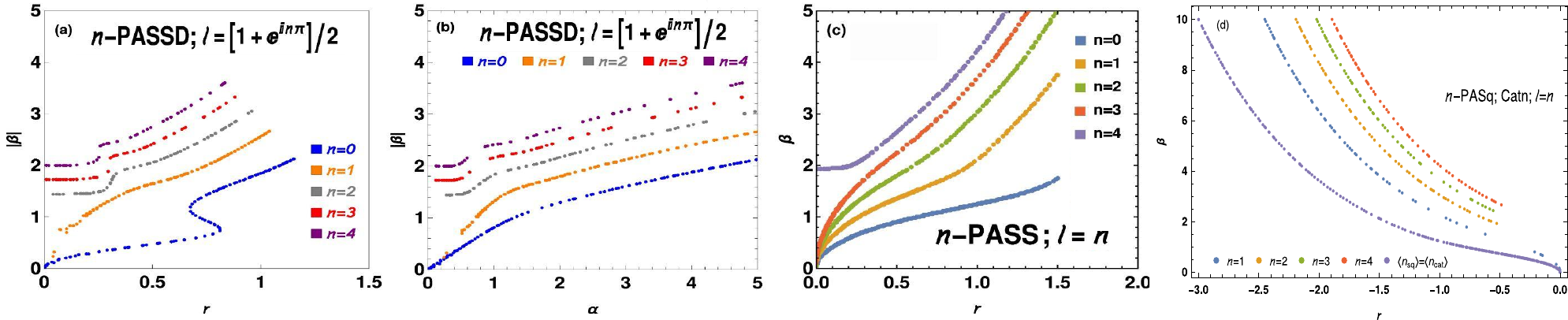}
    % \begin{minipage}[b]{0.31\linewidth}
    % \centering
    % \includegraphics[width=1.0\textwidth,height=.9\textwidth]{cntrssd01.pdf}
    % \end{minipage}
    % \hspace{0.05cm}
    % \begin{minipage}[b]{0.31\linewidth}
    % \centering
    %  \includegraphics[width=1.0\textwidth,height=.9\textwidth]{cntrssd02.pdf}
    % \end{minipage}
    % \hspace{0.05cm}
    % \begin{minipage}[b]{0.31\linewidth}
    % \centering
    %  \includegraphics[width=1.0\textwidth,height=.9\textwidth]{cntrss20.pdf}
    % \end{minipage}
    \caption{(Color online) Equal QFI (\(F_Q^\psi=F_Q^\phi\)) regimes are shown for \(n\)-PASSD vs. \(\psi_{\text{KS}}^l(-)\) in (a-b), \(n\)-PASS vs. \(\psi_{ks}^l(+)\) in (c), and \(n\)-PACatE vs. \(\psi_{\text{Cat}}^n\) in (d) across relevant parameter spaces. The parameters \(\{r,\alpha\}\in n-\)PASSD, while \(\beta\in\psi_{\text{KS}}^l(-)\) in plots (a) and (b); \(r\in n\)-PASS, and \(\beta\in\psi_{\text{KS}}^l(+)\) in (c); and \(r\in n\)-PACatE, and \(\beta\in\psi_{\text{Cat}}^l\) in (d). The plots (a-b) are 2D projections of the 3D contours satisfying \(F_Q^{n-\text{PASSD}} = F_Q^{\psi_{\text{KS}}^l(-)}\). The scattered data points form curves, indicating an increase in the target's amplitude (\(\beta\)) with each successive photon addition $n$.}
\label{numdist}
\end{figure}
\end{center}

In Fig.~\ref{numdist}(a-b), we plot the loci of equal Fisher information ($F_Q^{\text{proposed}} = F_Q^{\text{target}}$) for pair \textbf{prstrg-\#1}, parametrized by $(r, |\beta|)$ and $(\alpha, |\beta|)$, respectively. Similarly, Fig.~\ref{numdist}(c) and (d) displays the same constraint for \textbf{prstrg-\#2} and \textbf{prstrg-\#3}, in the space of $(r, |\beta|)$. The squeezing parameter 
$r$ applies to the SSD, SS, and Sq states, whereas the displacement parameters 
$\alpha$ and $\beta$ correspond to the SSD and the KS/Cat states, respectively. We assume $l=\left[1+e^{in\pi}\right]/2$ for \textbf{prstrg-\#1} and $l=n$ for both \textbf{prstrg-\#2}, and \textbf{prstrg-\#3}.
 
\begin{center}
\begin{figure}[htbp]
    \centering
     \includegraphics[height=0.28\linewidth, width=\linewidth]{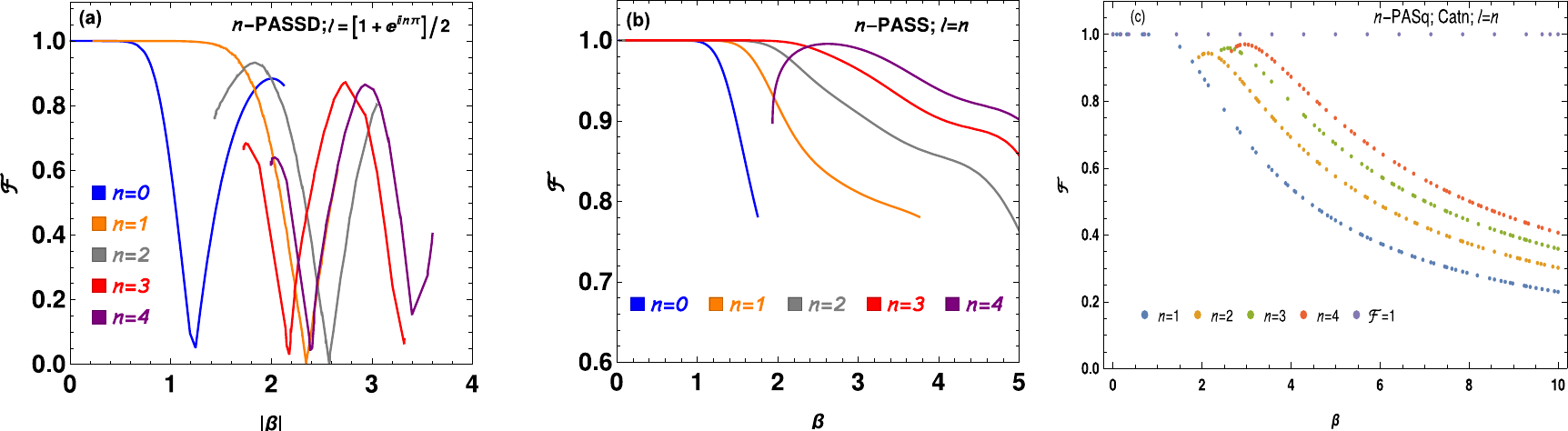}
    \caption{\justifying (Color online) The effect of photon additions on proposed SSD, SS and Sq, is clearly seen in high fidelity ($\mathcal{F}$), leading to an increase in $\beta$ of KS  in the plots (a) and (b), and of CatE in (c). (a) Fidelity remains close to one for $1$-PASSD against $\psi^{l}_{\text{KS}}(-)$ for $\beta$ less than 1.8, while the addition of more than 1 photon leads to a decrease in the maximum of fidelity to 0.9 around $\beta=2$. (b) Each subsequent PA to SS, leads to an increase in $\beta$ for fidelity$\,\sim\,1$. In (a-b), the $\{r,\alpha\}$ parameters satisfy equal Fisher information for states: ($n-$PASSD, $\psi^{l}_{\text{KS}}(-)$) and ($n-$PASS, $\psi^{l}_{\text{KS}}(+)$). Specifically, in plot (a), $\psi^{l}_{\text{KS}}(-)$ is parametrized with $l=(1+e^{in\pi})/2$, while in plot (b), $\psi^{l}_{\text{KS}}(+)$ takes $l=n$ as same as in (c) b/w $n-$PASq and Catn.}
    \label{fig:fid}
\end{figure}
\end{center}

These contours consistently show that achieving equal $F_Q$ leads to increasing the amplitude $|\beta|$ in KSs and CatE ($\psi^{0}_{Cat}$) as photon additions ($n$) on SS, SSD, and Sq increase. However, despite matching $F_Q$, the fidelity (evident from Fig.~\ref{fig:fid}) between KSs and $n-$PASS/SSD (or CatE and $n-$PASq) may remain low for large $\beta$ along contours seen in Fig.~\ref{numdist}(a-d), highlighting their distinct structures. This enables substitution of KSs by $n-$PASS or $n-$PASSD (or CatE by $n-$PASq) in high-precision small-displacement metrology. Conversely, $n$-PASSD and $n$-PASS, with high fidelity to KSs, are suitable for quantum error correction due to their resemblance to annihilation operator eigenstates (e.g., $\hat{a}^2 \ket{\text{cat}} = \alpha^2 \ket{\text{cat}}$, $\hat{a}^4 \ket{\text{KS}} = \alpha^4 \ket{\text{KS}}$~\cite{cat_code1}). To visualize these results, we next perform fidelity analysis for each case \textbf{prstrg-\#1}, \textbf{prstrg-\#2}, and \textbf{prstrg-\#3}.

\onecolumngrid

\twocolumngrid

\textbf{Fidelity ($\mathcal{F}$):} It is a widely used metric for quantifying the distinguishability between two quantum states $\ket{\psi}$ and $\ket{\phi}$ belonging to the same Hilbert space~\cite{cavesgeometry, zhou2019exact, armancompass}. For pure states, it is defined as the squared modulus of their overlap, expressed via the $\mathcal{L}^{2}$-norm:
\[
\mathcal{F} = \left| \langle \phi | \psi \rangle \right|^{2} = \left| \int_{-\infty}^{\infty} \psi(x) \phi^{*}(x)\, dx \right|^{2}.
\]
This formulation is valid for pure-state comparisons and is adopted throughout this work. Fidelity provides a robust measure of how closely a given quantum state approximates a target state. In our analysis, it is used to evaluate the proximity between proposed probe states and their corresponding reference or target states.

We observe that the fidelity measure ($\mathcal{F}$) remains close to unity for $0 < |\beta| < 1.8$ when $n = 1$, compared to the $n = 0$ case in Fig.~\ref{fig:fid}(a) for \textbf{prstrg-\#1}. For higher photon additions $n = 2$, $3$, and $4$, the fidelity decreases more noticeably; however, the overlap still remains above $80\%$. Figure~\ref{fig:fid}(a) compares the $n$-PASSD states with the target state $\psi_{KS}^{l}(-)$, while Fig.~\ref{fig:fid}(b) presents the analogous comparison for $n$-PASS states with $\psi_{KS}^{l}(+)$ in \textbf{prstrg-\#2}. Notably, in Fig.~\ref{fig:fid}(b), all cases with $n \geq 0$ exhibit high fidelity approaching unity at large KS output amplitude ($\beta$), which increases with higher photon additions. In the rightmost Fig.~\ref{fig:fid}(c), similar successive PAs lead to enhancing $\beta$ of CatE in \textbf{prstrg-\#3} with large $\mathcal{F}$ regimes close to one. This consistent trend of large overlaps and growing output amplitude is evident across both all subfigures~\ref{fig:fid}.

\onecolumngrid
\twocolumngrid

\section{conclusion}\label{sec:VII}
Non-Gaussian operations, particularly photon additions, are studied on various non-classical states such as compass and cat states. The effects of these repetitive operations on the states via sensitivity metric-QFI against phase-space shift $(s)$-leads to the fidelity as high as close to $1$, indicating the importance of these operations. We consider two different combinations of squeezing and displacement operators with opposite phase-space orientation. The parameter space of squeezing ($r$), displacement ($\alpha$) of proposed states, and $\beta$ amplitude of the KS and CatE, satisfies the same QFI, $F_{Q}^{\text{proposed}}=F_{Q}^{\text{Target}}$, against $s$. Interestingly, this parameter locus corresponds to the same average energy for the competing set of states with constraint for Fisher Information as $F_{Q}^{\psi}\propto \Delta^{2}\hat{x}(\theta)$ variance is a single-valued function of either squeezing and displacements. Finally, analysis of this locus reveals a region of high fidelity, leading to the state of high similarity and an increase in the amplitude ($\beta$) within same energy resource.

However, the realization of such non-Gaussian operations has been widely discussed through probabilistic models such as OPA and JC interactions in various platforms~\cite{chen2023generation}, it is important to analyze high-probabilistic models or approximate deterministic preparation of these photon-addition operations. Our proposed states with photon addition (i) $\hat{a}^{\dagger\,n}S[r](D[\alpha]+D[-\alpha])\ket{0}$, (ii) $\hat{a}^{\dagger\,n}(S[r]+S[-r])\ket{0}$ and (iii) $\hat{a}^{\dagger\,n}S[r]\ket{0}$, require availability of squeezing strength ($r$) and coherent amplitude ($\alpha$) to extent, leading to large amplitude. From previous work as well as these results of photon addition, conclude that large strengths of input parameters with small excitation number $n$ and vice-versa have regions of large fidelity. In preparation for proposed states, either through an exact Hamiltonian in trapped ions and superconducting circuits or via help of OPO, OPA and BS in optical platforms, keeping in mind difficulties involved in approaching operations namely ($\hat{a}^{\pm}$), we believe operations $(S[r]\pm S[-r])$ and $(D[\alpha]\pm D[-\alpha])$~\cite{Squeezed_hastrup,hastrupnature} could help realize previously discussed non-Gaussianity in limit of small $\alpha$ and $r$. This approximate relation can be viewed as 
$$    \lim_{r \to 0}
(S[r]-S[-r])^{n}\sim \left(\frac{\tanh(r)}{\sqrt{2}}\right)^{n}\hat{a}^{\dagger\,2n}$$ and, $$ \lim_{\alpha \to 0}
(D[\alpha]-D[-\alpha])^{n}\sim e^{-\frac{|\alpha|^{2}}{2}}\left(\alpha\right)^{n}\hat{a}^{\dagger\,n}.$$
 These non-unitaries can be implemented through dispersive interaction, as available in the known above discussed platforms, given their small interaction strength being related to the success probability of obtaining operations and introducing a relative phase between $S[r]$ and $D[\alpha]$.
Furthermore, such operations are discussed in superconducting circuits and trapped ions in the context of two-photon Rabi models and one-photon Rabi models with two-photon drive~\cite{armancompass}, leading to KSs.

\begin{acknowledgements}
Arman is thankful to the University Grants Commission and Council of Scientific and Industrial Research, New Delhi, Government of India for Junior Research Fellowship at IISER Kolkata.
\end{acknowledgements}

\appendix
\section{Notation}\label{appendix:a}

We provide a table for the notations used in the paper for various states and quantities for the sake of the reader.

%\begin{table}[h]
%\centering
%\begin{tabular}{|c|c|c|}
%\hline
%Notation & State & Other Notation \\
%$[$Acronm$]$ & $\ket{\psi}$ & [Math Symbol] \\
%\hline
%KS & $\sum\limits_{k=0}^{3}i^{-kl}\bar{f}_{k,k_{o}}\ket{i^{k}\alpha}$ & $\psi_{ks}^{l}(\pm)$ \\
%\hline
%$n$-PAKS & $\hat{a}^{\dagger\,n}\sum\limits_{k=0}^{3}i^{-kl}\bar{f}_{k,k_{o}}\ket{i^{k}\alpha}$ & $\hat{a}^{\dagger\,n}\psi_{ks}^{l}(\pm)$ \\
%\hline
%$n$-PAKSp & $\hat{a}^{\dagger\,n}\sum\limits_{k=0}^{3}i^{-kl}\bar{f}_{k,0}\ket{i^{k}\alpha}$ & $\hat{a}^{\dagger\,n}\psi_{ks}^{l}(+)$ \\
%\hline
%$n$-PAKSn & $\hat{a}^{\dagger\,n}\sum\limits_{k=0}^{3}i^{-kl}\bar{f}_{k,1}\ket{i^{k}\alpha}$ & $\hat{a}^{\dagger\,n}\psi_{ks}^{l}(-)$ \\
%\hline
%SSD & $\sum\limits_{k=0}^{1}S[r]\ket{e^{ik\pi}\alpha}$ & - \\
%\hline
%$n$-PASSD & $\hat{a}^{\dagger\,n}\sum\limits_{k=0}^{1}S[r]\ket{e^{ik\pi}\alpha}$ & - \\
%\hline
%SS & $\sum\limits_{k=0}^{1}S[e^{ik\pi}r]\ket{0}$ & - \\
%\hline
%$n$-PASS & $\hat{a}^{\dagger\,n}\sum\limits_{k=0}^{1}S[e^{ik\pi}r]\ket{0}$ & - \\
%\hline
%CatE & $\ket{\alpha}+\ket{-\alpha}$ & $\ket{\psi_{\textnormal{cat}}}$ \\
%\hline
%$n$-PACatE & $\hat{a}^{\dagger\,n}\ket{\psi_{\textnormal{cat}}}$ &$\hat{a}^{\dagger\,n}\ket{\psi_{\textnormal{cat}}}$ \\
%\hline
%$n$-PASq & $\hat{a}^{\dagger\,n}S[r]\ket{0}$ &$\hat{a}^{\dagger\,n}S[r]\ket{0}$ \\
%\hline
%\end{tabular}
%\caption{Nomenclature and notation used in the main text}
%\end{table}

\begin{table}[H]
\caption{\label{tab:nomenclature}Notation used in the main text, addressing their label and states' form.}
\centering
\begin{ruledtabular}
\begin{tabular}{lcl}
Notation & State $\ket{\psi}$ & Alternate Notation \\
\hline
KS/ks & $\displaystyle \sum_{k=0}^{3} i^{-kl} \bar{\mathsf{f}}_{k,k_{o}} \ket{i^{k}\alpha}$ & $\psi_{\text{ks}}^{l}(\pm)$ \\[3pt]

$n$-PAKS & $\displaystyle \hat{a}^{\dagger n} \sum_{k=0}^{3} i^{-kl} \bar{\mathsf{f}}_{k,k_{o}} \ket{i^{k}\alpha}$ & $\hat{a}^{\dagger n} \psi_{\text{ks}}^{l}(\pm)$ \\[3pt]

$n$-PAKSp & $\displaystyle \hat{a}^{\dagger n} \sum_{k=0}^{3} i^{-kl} \bar{\mathsf{f}}_{k,0} \ket{i^{k}\alpha}$ & $\hat{a}^{\dagger n} \psi_{\text{ks}}^{l}(+)$ \\[3pt]

$n$-PAKSn & $\displaystyle \hat{a}^{\dagger n} \sum_{k=0}^{3} i^{-kl} \bar{\mathsf{f}}_{k,1} \ket{i^{k}\alpha}$ & $\hat{a}^{\dagger n} \psi_{\text{ks}}^{l}(-)$ \\[3pt]

SSD & $\displaystyle \sum_{k=0}^{1} S[r]\ket{e^{ik\pi}\alpha}$ & -- \\[3pt]

$n$-PASSD & $\displaystyle \hat{a}^{\dagger n} \sum_{k=0}^{1} S[r]\ket{e^{ik\pi}\alpha}$ & -- \\[3pt]

SS & $\displaystyle \sum_{k=0}^{1} S[e^{ik\pi}r]\ket{0}$ & -- \\[3pt]

$n$-PASS & $\displaystyle \hat{a}^{\dagger n} \sum_{k=0}^{1} S[e^{ik\pi}r]\ket{0}$ & -- \\[3pt]

CatE & $\displaystyle \ket{\alpha} + \ket{-\alpha}$ & $\ket{\psi_{\text{cat}}}$ \\[3pt]

$n$-PACatE & $\displaystyle \hat{a}^{\dagger n} \ket{\psi_{\text{cat}}}$ & $\hat{a}^{\dagger n} \ket{\psi_{\text{cat}}}$ \\[3pt]

$n$-PASq & $\displaystyle \hat{a}^{\dagger n} S[r] \ket{0}$ & $\hat{a}^{\dagger n} S[r] \ket{0}$ \\
\end{tabular}
\end{ruledtabular}
\end{table}

\section{Quantum Fisher Information for States in the Main Text}\label{appendix:b}

%%%%%%%%%%%%%%%%%%%%%%%%%%%%%%%%5
We evaluate the quantum Fisher information (QFI) $F_Q^{\psi}$ for $n$-photon–added states using the generator $\hat{G}$ for phase-space displacement. For a small displacement $s = |s|e^{i\theta}$, the unitary $U = D[s] = e^{s\hat{a}^\dagger - s^*\hat{a}}$ leads to the generator
\[
G(\theta) = i\,U^\dagger\,\partial_{|s|}U\big|_{s=0} = \hat{a}^\dagger e^{i\theta} + \hat{a} e^{-i\theta}.
\]
For a pure state $\ket{\psi}$, QFI reduces to the variance of $\hat{G}$~\cite{liuquantum,Agerwal-Fisher}: $F_Q^\psi = \Delta^2 \hat{G}(\theta)$.

%We consider four states: (i) $\ket{\psi_{n\text{-PASSD}}} = \hat{a}^{\dagger n} S[r](D[\alpha]+D[-\alpha])\ket{0}$, (ii) $\ket{\psi_{n\text{-PASS}}} = \hat{a}^{\dagger n}(S[r]+S[-r])\ket{0}$, (iii) $\psi_{ks}^{l}(-) = \ket{\beta e^{i\pi/4}} - (-1)^{-l} \ket{-\beta e^{i\pi/4}} + i^{-l} \ket{i\beta e^{i\pi/4}} - (-i)^{-l} \ket{-i\beta e^{i\pi/4}}$, and (iv) $\psi_{ks}^{l}(+) = \ket{\beta} + (-1)^{-l} \ket{-\beta} + i^{-l} \ket{i\beta} + (-i)^{-l} \ket{-i\beta}$.
%
%Comparison of KSs with their photon-added versions ($n$-PAKSn, $n$-PAKSp) shows that non-Gaussian operations enhance KS amplitude, supporting the goal of increasing $\beta$ via such operations.

The QFI requires $\langle \hat{G} \rangle$ and $\langle \hat{G}^2 \rangle$, given by
\[
\langle G(\theta)^2 \rangle = f[n,n]^{-1} \sum_{k=0}^{2} \binom{2}{k} f[n+k, n+2-k] - 1,\] \[
\langle G(\theta) \rangle = \frac{f[n+1,n] + f[n,n+1]}{f[n,n]},
\]
where, $f[n,m] = \mathrm{Tr}(\hat{a}^n \hat{a}^{\dagger m} \ket{\psi}\bra{\psi})$.

These expressions are used to evaluate $F_Q^\psi$ for all states, considered in pairs such as $\mathbf{\textbf{prstrg}-\#1}, \mathbf{\textbf{prstrg}-\#2},\mathbf{\textbf{prstrg}-\#3} \mathbf{\textbf{trgtrgn}-\#1}$ and $\mathbf{\textbf{prstrgp}-\#2}$.

\begin{table}[H]
\caption{\label{tab:fnm_expressions}
Expressions for the function $f[n, m]$ corresponding to PA variants with $\mathcal{C}_{n,m}^{r_1,r_2}(\alpha,\beta)$ denotes a cross term squeezing and displacement parameters.}
\centering
\begin{ruledtabular}
\begin{tabular}{lc}
State $\psi$ & $f[n,m]$ \\
\hline
$n_{o}$-PASSD &
$\displaystyle \sum_{k,l=0}^{1}
    e^{i\theta (m-n)}\mathcal{C}_{n,m}^{-r,r}(\alpha e^{ik\pi},\,\alpha e^{il\pi})$ \\[3pt]

$n_{o}$-PASS &
$\displaystyle \sum_{k,l=0}^{1}
    e^{i\theta (m-n)}\mathcal{C}_{n,m}^{(-1)^{k}r,(-1)^{l}r}(0,\,0)$ \\[3pt]

$n_{o}$-PAKS &
$\displaystyle \sum_{k,l=0}^{3}
    e^{i\theta (m-n)\pi}i^{(l-k)l_{o}}\bar{\mathsf{f}}_{k,k_{o}}\bar{\mathsf{f}}^*_{l,k_{o}}\mathcal{C}_{n,m}^{0,0}(i^{l}\alpha,\,i^{k}\alpha)$ \\
\end{tabular}
\end{ruledtabular}
\end{table}

\begin{table}[H]
\caption{\label{tab:f_bar_values}
Values of $\bar{\mathsf{f}}_{k_1, k_0}$ used in defining the KSs $\psi_{KS}^{l_0}(+)$ and $\psi_{KS}^{l_0}(-)$ and PAKSs, where $k_0 = 0$ denotes the even KS and $k_0 = 1$ the odd KS.
}
\centering
\begin{ruledtabular}
\begin{tabular}{ccc}
$k_1$ & $k_0$ & $\bar{\mathsf{f}}_{k_1,k_0}$ \\
\hline
$0$ & $[0,1]$ & $[1,1]$ \\
$1$ & $[0,1]$ & $[1,1]$ \\
$2$ & $[0,1]$ & $[1,-1]$ \\
$3$ & $[0,1]$ & $[1,-1]$ \\
\end{tabular}
\end{ruledtabular}
\end{table}

% \[
% \begin{array}{|c|c|}
% \hline
% \text{State} (\psi) & f[n,m]\\
% \hline
% n_{o}\text{-PASSD} & \sum\limits_{k,l=0}^{1}
%     e^{i\theta (m-n)}\mathcal{C}_{n,m}^{-r,r}(\alpha e^{ik\pi},\,\alpha e^{il\pi}) \\
% \hline
% n_{o}\text{-PASS} & \sum\limits_{k,l=0}^{1}
%     e^{i\theta (m-n)}\mathcal{C}_{n,m}^{(-1)^{k}r,(-1)^{l}r}(0,\,0) \\
% \hline
% n_{o}\text{-PAKS} & \sum\limits_{k,l=0}^{3}
%     e^{i\theta (m-n)\pi}i^{(l-k)l_{o}}\bar{f}_{k,k_{o}}\bar{f}^*_{l,k_{o}}\mathcal{C}_{n,m}^{0,0}(i^{l}\alpha,\,i^{k}\alpha) \\
% \hline
% \end{array}
% \]

Here, cross term ($\mathcal{C}$) used in the above table for $n_{o}$ photon-added states, is
\begin{multline*}
     \mathcal{C}_{n,m}^{r_{1},r_{2}}(\alpha,\beta)= \bra{\alpha}S[r_{1}]a^{n}a^{+m}S[r_{2}]\ket{\beta} \\=\Tr[\hat{a}^{p_{o}}\hat{a}^{\dagger q_{o}}\hat{a}^{\dagger n_{o}}S[r_{2}]\ket{\beta}\bra{\alpha}S[r_{1}]\hat{a}^{n_{o}}]\\
     =\frac{(-1)^{n}}{\sqrt{\cosh[r]}}\partial_{\gamma^{*}}^{n}\partial_{\gamma}^{m}\left[e^{\tanh{[r]} (\frac{\delta^{2}}{2})-\frac{(|\delta|^{2}+|\gamma|^{2})}{2} }\right.\\\left.\left.e^{i\text{Im}[\beta^*(\bar{\gamma}-\bar{\alpha}_{r})-\bar{\gamma}^*\bar{\alpha}_{r}]}\right]\right|_{\gamma=0},
\end{multline*} with $r=r_1 +r_2,$ $\bar{\gamma}=\gamma \cosh{[r_{2}]} + \gamma^{*}\sinh{[r_{2}]},$ $\alpha_{r}=\alpha \cosh{[r]} + \alpha^{*}\sinh{[r]},$ $\delta = \beta + \bar{\gamma}-\bar{\alpha}_{r}$ and indices $(n,m) =(n_{o}+q_{o}, n_{o}+p_{o})$  are shuffled with ($q_{o}, p_{o}$) in the evaluation of the $\langle \hat{G}^{p}\rangle.$ 
Moreover, quantity \begin{multline*}
    \bar{\mathsf{f}}_{k_{1},k_{0}}=2^{\frac{k_{0}}{2}}\left[\sin{\left(\frac{(2k_{1}+1)\pi}{4}\right)}\right]^{k_{0}}\\=e^{\frac{1}{2} i \pi  k_{0} \left[-\sin \left(\frac{\pi k_{1}}{2}\right)+\cos \left(\frac{\pi k_{1}}{2}\right)+1\right]},
    \end{multline*}
denotes $\psi^{l_{o}}_{ks}(+)$ and its photon added variants for $k_{0}=0$, while it represents $\psi_{ks}^{l_{o}}(-)$ and its photon added variants for $k_{o}=1.$

\bibliography{biblio.bib}

\end{document}